\documentclass{IEEETran}

\usepackage{graphicx}
\usepackage{dcolumn}
\usepackage{bm}
\usepackage[utf8]{inputenc} 
\usepackage{amsmath} 
\usepackage{array}
\newcolumntype{M}[1]{>{\centering\arraybackslash}m{#1}}
\newcolumntype{C}[1]{>{\centering\arraybackslash}p{#1}}
\usepackage{graphicx, rotating}
\usepackage{subfig}
\usepackage{siunitx}
\usepackage{tabularx}
\usepackage{ulem}
\usepackage{enumitem}
\setitemize{noitemsep,topsep=0pt,parsep=0pt,partopsep=0pt}
\usepackage{wrapfig}
\usepackage[colorlinks = true,
            linkcolor = black,
            urlcolor  = black,
            citecolor = black,
            anchorcolor = black]{hyperref}
\usepackage{fancyhdr} 
\usepackage[left=3cm,right=2cm,top=2.5cm,bottom=2.5cm,includeheadfoot]{geometry}
\usepackage{lmodern}
\usepackage{chngcntr}
\usepackage{float}
\usepackage{xcolor}
\usepackage{wrapfig}
\usepackage{booktabs}
\usepackage{esvect}
\usepackage{diagbox}
\graphicspath{ {./images/} }

\usepackage[utf8]{inputenc}
\usepackage[T1]{fontenc}

\newcommand{\ds}{\ensuremath{\displaystyle}}

\begin{document}

\title{Design of an electrostatic balance mechanism to measure optical power of \SI{100}{\kilo\watt}}

\author{\IEEEauthorblockN{Lorenz Keck\IEEEauthorrefmark{1},
Gordon Shaw\IEEEauthorrefmark{1},
Ren\'e Theska\IEEEauthorrefmark{4}, 
Stephan Schlamminger\IEEEauthorrefmark{1}} 
\IEEEauthorblockA{\IEEEauthorrefmark{1}Physical Measurement Laboratory, National Institute of Standards and Technology,
Gaithersburg,  MD 20899 USA}
\IEEEauthorblockA{\IEEEauthorrefmark{4} Precision Engineering Group, Technische Universit\"at Ilmenau, 98693 Ilmenau, Germany} 
\thanks{Manuscript received December 1, 2020;  
Corresponding author: S. Schlamminger (email: stephan.schlamminger@nist.gov)}}

\date{\today}

\IEEEtitleabstractindextext{%
\begin{abstract}
A new instrument is required to accommodate the need for increased portability and accuracy in laser power measurement above $\SI{100}{\watt}$. Reflection and absorption of laser light provide a measurable force from photon momentum exchange that is directly proportional to laser power, which can be measured with an electrostatic balance traceable to the SI. We aim for a relative uncertainty of $10^{-3}$ with coverage factor $k=2$. 
For this purpose, we have designed a monolithic parallelogram 4-bar linkage incorporating elastic circular notch flexure hinges. The design is optimized to address the main factors driving force measurement uncertainty from the balance mechanism: corner loading errors, balance stiffness, stress in the flexure hinges, sensitivity to vibration, and sensitivity to thermal gradients. Parasitic rotations in the free end of the 4-bar linkage during arcuate motion are constrained by machining tolerances. An analytical model shows this affects the force measurement less than 0.01 percent. Incorporating an inverted pendulum  reduces the stiffness of the system without unduly increasing tilt sensitivity.  Finite element modeling of the flexures is used to determine the hinge orientation that minimizes stress which is therefore expected to minimize hysteresis. Thermal effects are mitigated using an external enclosure to minimize temperature gradients, although a quantitative analysis of this effect is not carried out. These analyses show the optimized mechanism is expected to contribute less than \SI{1e-3}{} relative uncertainty in the final laser power measurement.
\end{abstract}

}

\maketitle
\IEEEdisplaynontitleabstractindextext

\section{Introduction}

Primary measurements of laser power rely on either the effect of absorbed laser power or the force transmitted in reflection, see~\cite{Williams.2020} for a recent review.
Instruments that use absorption suffer three critical disadvantages. 
First and foremost, the laser beam is no longer available after it has been absorbed. This necessitates substitution or beam splitting processes for calibration of secondary detectors and severely limits in-situ use for industrial applications such as, e. g., laser welding.
Second, every absorber scatters and reflects some light.
Hence, it is difficult to capture the entirety of the incident light at relative uncertainties smaller than $\SI{1e-3}{}$. This is expected to be especially critical for laser powers above \SI{1}{\kilo\watt}. 
Third, absorption of multiple kilowatts of laser power generates a large amount of heat. Although flowing water calorimeter systems capable of handling these thermal loads have been developed~\cite{Williams.2020} they are bulky and difficult to operate.

In contrast to that, custom dielectric coating stacks that have total optical loss lower than $\SI{1e-4}{}$ are commercially available. Thus it is, in principle, possible to build a system that can measure the power of a multi-kilowatt laser at relative uncertainty of $\SI{1e-3}{}$ or better with $k=2$, using the photon pressure force from reflection of laser light.

The optical characteristics of the mirror can be described by the specular reflectance $R$, the absorbance $A$, and the transmittance $T$. The effect of diffuse reflectance is not considered in this work. Their sum is unity, i. e., $R+A+T=1$. Using these coefficients~\cite{Shaw.2019b}, the photon pressure force, throughout the text also referred to as the external force, on the mirror is given by 
\begin{equation}		
F_\mathrm{ext} =  \frac{P \cos{\alpha}}{c} \left(2 R+A \right)\label{eq:photontrans2}.
\end{equation}
Here, $P$ denotes the power of the laser beam and $\alpha$ the angle of incidence relative to surface normal. 
A \SI{100}{\kilo \watt} light beam ($\alpha=0$) normally incident on a perfect mirror ($R=1$) produces a force of $\SI{667}{\micro\newton}$. According to the specifications above, the total allowable force uncertainty  is  $\SI{667}{\nano\newton}$ at $k=2$.
Since we do not have an \SI{100}{\kilo \watt} laser at our disposal, a beam multiplier, the High Amplification Laser-pressure Optic (HALO) has been constructed~\cite{AlexandraArtusioGlimpseKyleRogersPaulWilliamsandJohnLehman.}. The HALO uses a  \SI{10}{\kilo \watt} laser and 14 reflections to produce a normal force on the order of $\SI{667}{\micro\newton}$. Other work describes multi-reflection measurements at lower power~\cite{Shaw.2019,Vasilyan.2020}.  In this article, we describe  the design of the mechanical components of an electrostatic balance for measuring laser power of up to \SI{100}{\kilo\watt}. We designed the  electrostatic balance to be compatible with the HALO, but it can also be used to measure a single laser beam application. 
This manuscript reuses some content from thesis~\cite{Keck.2020} with permission.

\section{Theory of the electrostatic balance}

Electrostatic force balances have been used successfully in mass metrology~\cite{Shaw.2019, Shaw.2016}, and, more recently, to measure the force exerted by  light for power levels up to \SI{3}{\watt}~\cite{Shaw.2019b}.

In force mode, an external force is compensated by the electrostatic force between two capacitor electrodes. 
A feedback system adjusts the voltage applied to the capacitor to hold a movable electrode at a nominal position based on input from a displacement measuring device, usually an interferometer.
The electrostatic force generated with the capacitive actuator depends on two quantities; the square of the potential difference, $V^2$, and the capacitance gradient $\mathrm{d}C/\mathrm{d}x$, and is described with
\begin{equation}
F_{\mathrm{el}} = -\frac{1}{2} \frac{\mbox{d}C}{\mbox{d}x} V^2.\label{eq:efb}
\end{equation}

The capacitance gradient, $\mbox{d}C/\mbox{d}x$ is not known a priori.
It is obtained by measuring capacitance at fixed electrode positions, fitting a polynomial to the measured data and then calculating the derivative of the polynomial about the nominal operating position. 
Excursions of a few tenths of a millimeter are required to measure the capacitance gradient with sufficient accuracy~\cite{Stirling.2017}. 
Gradients on the order of $\SI{1}{\pico \farad \per \milli \meter}$ are achieved with concentric cylindrical capacitors as in electrostatic force balances at NIST~\cite{Shaw.2016, Stirling.2017}.

Measurements performed with the electrostatic balance are directly traceable to the SI, as revised in 2019~\cite{Stock_2019}. 
The measurement may therefore be considered a primary reference for force, as no other reference, i. e., a force traceable to mass in a gravitational field, is required for calibration. More details to the traceability path of measured force to $h$ are outlined in \cite{Shaw.2016} and \cite{Williams.2020}. 

\section{Conceptual design of a new balance mechanism}
 
\begin{figure}[htb]
    \centering
    \includegraphics[width=.26\textwidth]{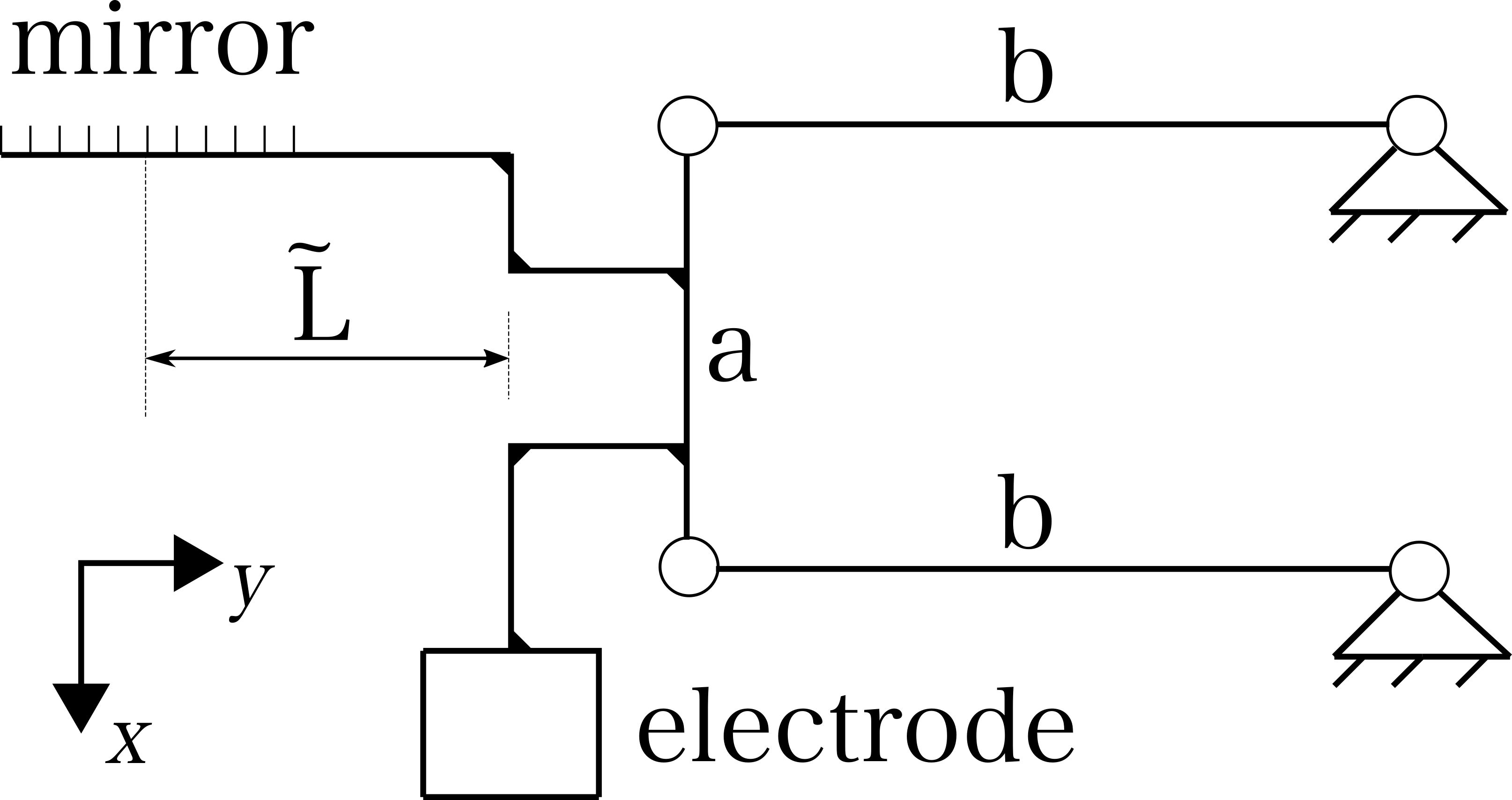}
    \caption{Drawing of the mechanism with attached inner capacitor electrode and mirror.  The coupler with length $a$ is shown on the left. The swings, with length $b$ determine the distance between the fixed back pivots and the movable front pivots. $\Tilde{L}$ indicates an offset of the center axis of the mirror to the center axis of the capacitor.}
    \label{fig:principle}
\end{figure}

The design objective is to achieve linear translation of the payload (mirror plus electrode) over $\pm\SI{0.25}{\milli\meter}$ with minimal parasitic rotation.
Furthermore, the mechanism should be as simple as possible for ease of use, manufacturing, and uncertainty analysis.
A planar parallelogram linkage, see Fig.~\ref{fig:principle}, is a suitable solution for this application.

The mechanism has four pivots. The two back pivots connect two rotating bars (referred to as swings) to the frame of the balance and the two front pivots connect the coupler to the rotation bars. Two dimensions must be chosen: the vertical separation of the pivots, i. e., the length of the coupler, $a$, and the length of the swings, $b$.

\subsection{Sizing the linkage}

As shown in Fig.~\ref{fig:principle} the  center  of the mirror is horizontally offset from the  center axis of the capacitor to prevent heating of sensitive components by transmitted light~\cite{Shaw.2019b}.   The variable $\Tilde{L}$ denotes the horizontal offset between the application points of the external force due to the laser and the electrostatic force.
This lever arm increases the measurement sensitivity to coupler rotations.
In an ideal parallelogram linkage, where the four pivots are at the corner of a perfect parallelogram, the coupler will not rotate. In reality, a perfect parallelogram is impossible to achieve  due to machining tolerances, $\Delta$. 

\begin{figure}[htb]
    \centering
    \includegraphics[width=0.8\columnwidth]{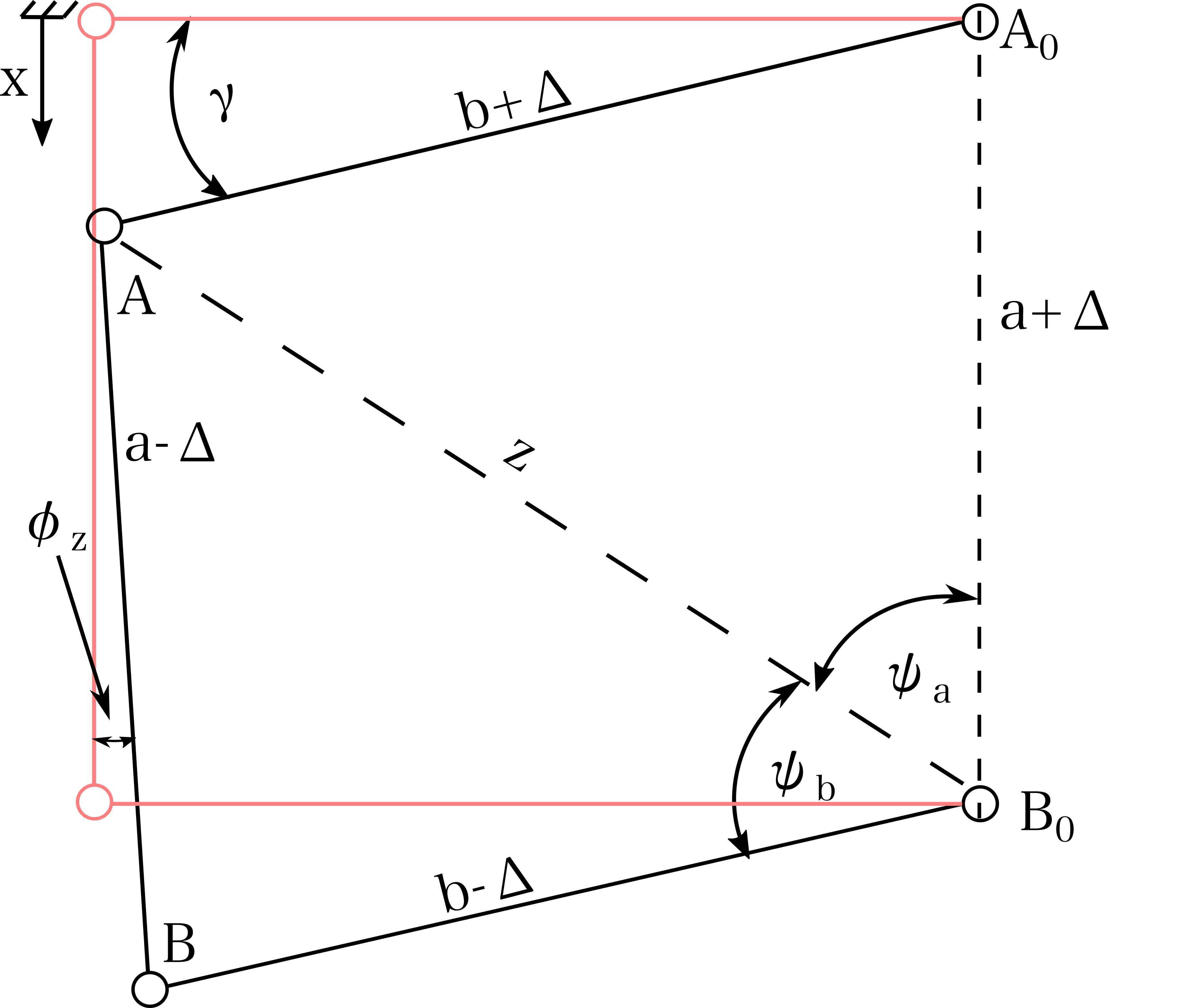}
    \caption{The geometric relations for an assumed imperfect parallelogram linkage with $b$ - length of the swings, $a$ - length of the coupler, $\Delta$ - manufacturing tolerances.  A worst case scenario is assumed:  the opposite linkages differ by $2\Delta$.  The dashed line  $z$  is one diagonal. Here, $\psi_\mathrm{a}, \psi_\mathrm{b}$ are calculated as function of $\gamma$.   The  parasitic rotation of the coupler is $\phi_\mathrm{z}$. A$_\mathrm{0}$ and B$_\mathrm{0}$ display fixed back pivots while A and B are moveable hinges. The red lines show the linkage at the nominal zero position and the black lines exaggerate a deflected state of the linkage.
\label{fig:zeta}}
\end{figure}

To examine the effect of machining tolerances on the rotation angle of the coupler, $\phi_\mathrm{z}$, a worst case is assumed. 
The horizontal distance at the top is $2\Delta$ longer than the one at the bottom. Further, the coupler length is $2\Delta$ longer than the vertical separation of the back pivots, see Fig.~\ref{fig:zeta}.

The squared length of the diagonal shown in Fig.~\ref{fig:zeta} is given by 
\begin{multline}
    z^2 = (a + \Delta )^2+(b + \Delta )^2 \\
    \hspace{6ex}-2   (a + \Delta )   (b + \Delta )   \sin{(\gamma)}, \hfill
\end{multline}
where the rotation angle of the top swing $\sin{(\gamma)} = x/(b + \Delta) \approx \gamma$.
With the length of the diagonal, the angles around point B$_\mathrm{0}$ can be obtained with the cosine rule. They are
\begin{equation}
    \psi_\mathrm{a} =\arccos{\left(\frac{(a + \Delta )^2+z^2-(b + \Delta )^2}{2   z   (a + \Delta )}\right)},
\end{equation}
and 
\begin{equation}
\psi_\mathrm{b} = \arccos{\left(\frac{(b - \Delta )^2 + z^2 - (a - \Delta )^2}{2   z   (b - \Delta )}\right)}.
\end{equation}

The rotation angle of the coupler is 
\begin{multline}
    \phi_\mathrm{z}(x) =\\ \arcsin{\left(\frac{(b + \Delta )   \cos{(\gamma)} -(b - \Delta )   \sin{(\psi_\mathrm{a}+\psi_\mathrm{b})}}{(a - \Delta )}\right)}.\label{eq:zeta}\\
\end{multline}    

To minimize coupler rotation the first derivative, $\mbox{d}\phi_\mathrm{z}/\mbox{d}x$, should be zero at the nominal zero position ($\gamma=0$).
This is the case for the perfect geometry, $\Delta =0$.

\begin{figure}[hbt]
    \centering
    \includegraphics[width=3.2in]{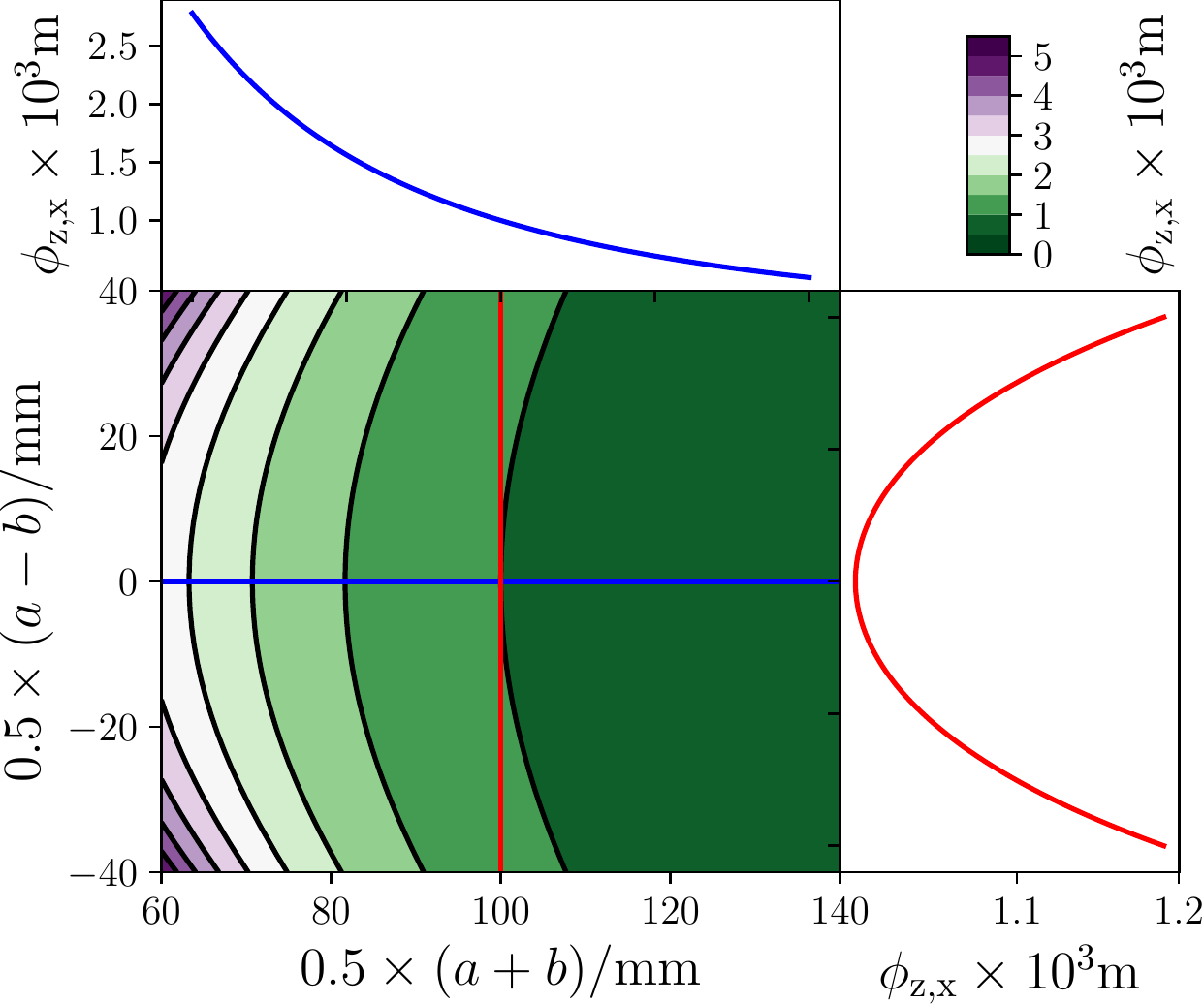}
    \caption{The figure of merit, $\phi_{\mathrm{z,x}}=\mbox{d} \phi_\mathrm{z}/\mbox{d} x$, for the optimization of the parallelogram sides $a$ and $b$. For this figure, $\Delta=\SI{5}{\micro \meter}$ is used.
\label{fig:phiz_a+b}}
\end{figure}

Hence, $\mbox{d}\phi_\mathrm{z}/\mbox{d}x$ is a good  figure of merit to investigate the coupler rotation.
Fig.~\ref{fig:phiz_a+b} shows the derivative as a function of the average $(a+b)/2$ and half the difference $(a-b)/2$ of two lengths. Here, $\Delta=\SI{5}{\micro \meter}$ was used for the calculation -- uncertainties that can readily be achieved with computer numerically controlled  machining.
For any average value chosen, the figure of merit shows a minimum for $a=b$. Hence, $a=b$ is a good choice for the  design. 

Since the coupler rotation  decreases monotonically
 as $a+b$ increases
, the requirement for a compact instrument dictates the choice of \SI{100}{\milli\meter} for $a$ and $b$.

For the chosen geometry a coupler rotation of $\mbox{d}\phi_\mathrm{z}/\mbox{d}x= \SI{1}{\micro \radian \per \milli \meter}$ is obtained.
The distance to the apparent center of rotation is the reciprocal of this value, $ L = (\mbox{d} \phi_\mathrm{z}/\mbox{d} x)^{-1}= \SI{1}{\kilo\meter}$.

$L$ constrains the systematic and statistical uncertainty caused by the horizontal difference in the application points of the external force and the compensating force, $\Tilde{L}$ in Fig.~\ref{fig:principle}.
 With respect to the rotation point, the electrostatic force $F_\mathrm{el}$ produces a torque of $ L\cdot F_\mathrm{el}$, while the 
external force $F_\mathrm{ext}$  produces a torque $(L\pm\tilde{L}) \cdot F_\mathrm{ext}$. The sign is positive if the rotation point is to the right of the coupler. In equilibrium, both torques must be equal and, hence, the relative force difference is  $F_\mathrm{el}/F_\mathrm{ext}-1=\pm\tilde{L}/L $.
This expression describes the corner loading error.

With typical values of $\Tilde{L}=\SI{90}{\milli \meter} \pm \SI{2}{\milli \meter}$ and $L = \SI{1}{\kilo\meter}$, the relative corner loading error is $\SI{9e-5}{}\pm \SI{2e-6}{}$ which shows that this error contribution is orders of magnitude below the required \SI{1e-3}{}.

\subsection{Analytical description}\label{subsec:analytical}

\begin{figure}[htb]
    \centering
    \includegraphics[width=1.0\columnwidth]{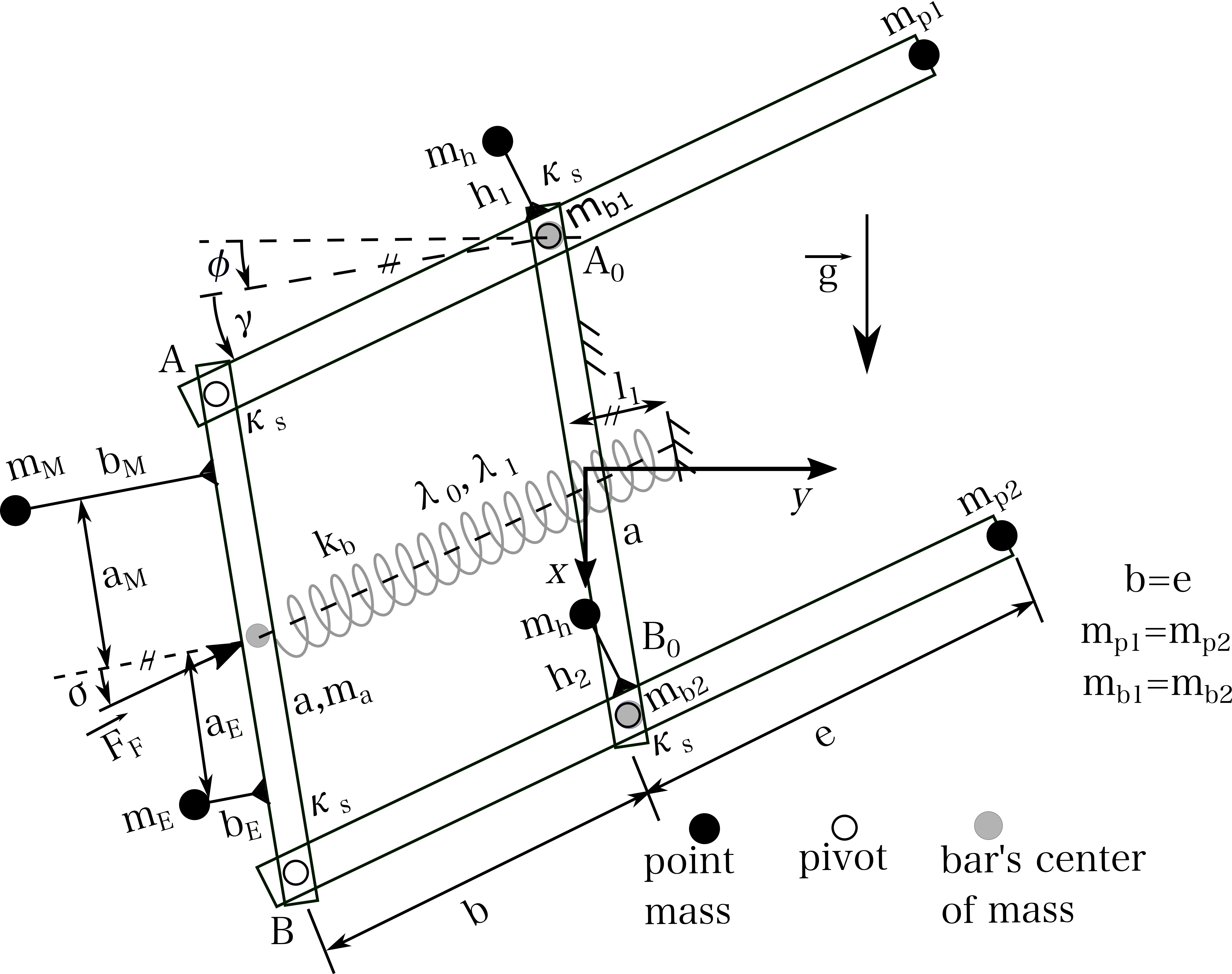}
    \caption{Rigid body  model of the mechanism with its attachments.}
    \label{fig:mechanismvertical}
\end{figure}

In this section, the balance is analyzed using the Lagrange equations of the second kind. 
This will yield  the stiffness, the eigenfrequency,  and the condition for the equilibrium position of the mechanism.
The functional components (masses, springs, and pivots) are shown in Fig.~\ref{fig:mechanismvertical}. 
All connecting bars are assumed to be perfectly rigid and all damping is neglected. 
A single hinge has a torsion stiffness of $\kappa_\mathrm{s}$,  as indicated by the subscript s for single. 
The $x$ axis of the coordinate system is aligned with gravity, the metrology frame is inclined by $\phi$ from the $x$ axis, and the rotating links are deflected by $\gamma$ from the metrology frame.
The two masses $m_{\mathrm{h}}$ are offset by
$h_{\mathrm{1}}$ and $h_{\mathrm{2}}$  along the negative $x$ direction from the back pivots in a non-deflected system, i. e., for $\gamma=\phi=0$.
These two masses $m_{\mathrm{h}}$ and the compensation spring labelled $k_\mathrm{b}$ can be used to  adjust the mechanism stiffness~\cite{Darnieder.2019, JonP.}. 
Besides the mechanical stiffness $k_\mathrm{b}$ the zero length $\lambda_0$ and the extended length $\lambda_1$ are the important physical parameters for the spring.
The masses $m_{\mathrm{p1}}$  and $m_{\mathrm{p2}}$ are counterweights and compensate  the masses $m_{\mathrm{a}}$ (coupler), $m_{\mathrm{M}}$ (mirror), and $m_{\mathrm{E}}$ (capacitor electrode). Here,  $a_\mathrm{E}$/$b_\mathrm{E}$ and  $a_\mathrm{M}$/$b_\mathrm{M}$  denote the vertical/horizontal distances from the center of the coupler to the electrode and the mirror, respectively. The symbols $a$ and $b$ without indices abbreviate the lengths  of the parallelogram, similar to Fig.~\ref{fig:zeta}. The symbol $e$ captures the length of the extension of the upper or lower swing to the right of the back pivots, to the counterweights.

The differential equation for $\gamma$ can be obtained from the Lagrange equation assuming small angles  $\gamma$ and $\phi$. It is derived in appendix A and the result can be written in the following form,
\begin{equation}
    J \ddot{\gamma} + \kappa \gamma = -J_{\mathrm{\phi}} \ddot{\phi}  + \kappa_{\mathrm{\phi}} \phi - N_{\mathrm{eq}},\label{eq:eqm}
\end{equation}

where the coefficients are given by
\begin{multline}
    J = b^{2} m_\mathrm{E} + b^{2} m_\mathrm{M} + b^{2} m_\mathrm{a} + \frac{b^{2} m_\mathrm{b}}{6} + \frac{b e m_\mathrm{b}}{3} + \frac{e^{2} m_\mathrm{b}}{6} \\
    \hspace{6ex}  +2 e^{2} m_\mathrm{p} + h_\mathrm{1}^{2} m_\mathrm{h} + h_\mathrm{2}^{2} m_\mathrm{h}, \hfill \label{eq:J_eq}
\end{multline}

\begin{multline}
J_{\mathrm{\phi}} = J +\frac{a h_\mathrm{1} m_\mathrm{h}}{2} - \frac{a h_\mathrm{2} m_\mathrm{h}}{2} + b b_\mathrm{e} m_\mathrm{E} + b b_\mathrm{m} m_\mathrm{M}  \hfill 
\end{multline}

\begin{multline}
    \kappa = 4 \kappa_\mathrm{s} - g m_\mathrm{h} \left(h_\mathrm{1} + h_\mathrm{2}\right) \\
    \hspace{6ex} - l_{\mathrm{1}} b k_\mathrm{b} \left(1 - \frac{\lambda_{\mathrm{0}}}{b+l_{\mathrm{1}}}\right), \hfill \label{eq:K_eq}
\end{multline}
\begin{flalign}
     \hspace{2.5ex}&\kappa_{\mathrm{\phi}} = g m_{h} \left(h_\mathrm{1} + h_\mathrm{2}\right) \label{eq:K_eq,phi},\;\mbox{and}& \hfill
\end{flalign}
\begin{flalign}
    \hspace{2.5ex}&N_{\mathrm{eq}} = g \left(- b m_\mathrm{E} - b m_\mathrm{M} - b m_\mathrm{a} + 2 e m_\mathrm{p}\right).& \hfill \label{eq:N_eq}
\end{flalign}

The imaginary eigenvalues of the homogeneous part of the differential equation provide the eigenfrequency $\omega$ due to deflections $\gamma$. It is
\begin{equation}
\omega^2 = \frac{\ds \kappa}{\ds  J}.
\end{equation}
The torque $N_{\mathrm{eq}}$ determines the equilibrium position according to $\gamma= N_{\mathrm{eq}}/\kappa$.  
The nominal zero position ($\gamma=0$ for $\phi=0$) can be obtained according to the equilibrium condition in Eq.~\ref{eq:N_eq} by adjusting the counterweights such that $2 e m_\mathrm{p} =b (m_\mathrm{E}+ m_\mathrm{M} + m_\mathrm{a})$. 
Then, the load is distributed equally to the upper and lower pivots.
The masses on both sides of the $xz$ plane generate equal and opposite torques, minimizing the effects of external vertical acceleration.
By choosing $b=e$, the equilibrium position of the balance remains largely unchanged with temperature change, 
since thermal expansion in both lengths $b$ and $e$ would cause the lever arms to expand symmetrically, and thus the equilibrium condition stays stable. Nevertheless, a temperature gradient within the material could lead to asymmetric thermal expansion, but since the chosen Aluminum alloy has a high thermal conductivity, this effect is considered negligible.\cite{JonP.}.

The torsional stiffness $\kappa$ given in Eq.~\ref{eq:K_eq} can be converted to a linear stiffness $K$ of the coupler moving in $x$ with~\cite{JonP.}:
\begin{equation}
    K = \frac{\kappa}{b^2}.\label{eq:transstiff}
\end{equation}
It can be seen that the stiffness decreases with $1/b^2$. 
Hence, to obtain the necessary small linear stiffness, $b$ should be as large as possible, but, as mentioned above, the installation space limits $b$ to a maximum of $\SI{100}{\milli\meter}$.

\subsection{Monolithic design}

A compliant mechanism is a key part for the realization of precision balance instrumentation. 
Flexure hinges need no lubrication, are stick-slip free, show negligible hysteresis and provide highly reproducible motion.~\cite{Howell.2013, Lobontiu.2003, Zentner.2014}

A monolithic design also has several advantages. 
Fabricating the functional parts of the mechanism in one setup maintains small machining tolerances yielding two major benefits: (1) The rotation axes of the four pivots are parallel, and (2) the machining tolerances $\Delta$ are small. Furthermore, no assembly is required, saving time and  eliminating a source of potential variation from the model. Hence, nearly identical copies can be made. The lack of fasteners also reduces excess mass.

Many different possible contours for hinges exist. A detailed overview can be found in~\cite{Lin.2015}. They differ in three functional properties, (1) the torsional stiffness, (2) the stability of the rotation axis under deflection, and (3) the maximum admissible deflection. A small rotational stiffness and a good stability of the axis of rotation are both important in the current design.  

The linear stiffness is given by a combination of Eq.~\ref{eq:K_eq} and Eq.~\ref{eq:transstiff}. Here, a stiffness of \SI{0.1}{\newton\per\meter} is desired. For the chosen $b=\SI{100}{\milli \meter}$ and no stiffness compensation, the rotational stiffness of one pivot should be $\kappa_{\mathrm s}=\SI{2.4e-4}{\newton \cdot \meter}$.
Generally, a circular geometry has moderate bending stiffness, is easy to manufacture, and the precision of rotation is high compared to other geometries ~\cite{Dirksen.2011}.  The latter is because the compliant part of the hinge performing the rotation is concentrated in the very center of the hinge geometry. 
For simplicity, the following circular hinge contour was chosen for each flexure hinge: radius \SI{2.5}{\milli \meter}, minimal notch height \SI{0.05}{\milli\meter} and width of \SI{10}{\milli \meter}. These dimensions can be obtained with high speed milling or wire electrical discharge machining.

The rotational stiffness of this hinge design is $\kappa_\mathrm{s}=\SI{0.018}{\newton \cdot \meter}$, as calculated with non-linear equations of large deflections assuming a pure moment loading a hinge~\cite{Henning.2018}.
This is about two orders of magnitude larger than desired, but the final stiffness of the mechanism can be adjusted with the masses $m_\mathrm{h}$. 
The final design keeps the inverted pendulum to reduce stiffness, but omits the spring to reduce temperature sensitivity which would have affected the elastic modulus
, and, hence, the equilibrium position of the balance.
The maximum admissible angle of deflection of the chosen hinge  under a pure moment load is $\approx\pm\SI{87}{\milli \radian}$ and provides more than an order of magnitude more than the required $\pm\SI{2.5}{\milli \radian}$ for a coupler travel of $\pm\SI{0.25}{\milli\meter}$ with the chosen linkage dimensions.
Aluminum 7075-T6 was chosen as the material for the monolithic mechanism due to its high yield strength ($\SI{503}{\mega\pascal}$), low elastic modulus ($\SI{72}{\giga\pascal}$), and good machineability. 

\section{Detailed design by use of finite element analysis}
\label{sec:detaildesign}

With the design analysis above as a starting point, the mechanism can be refined to its final form. This section describes an optimization for robustness, functionality, portability, and machinability using finite element methods.

\subsection{Optimization of hinge orientation}

\begin{figure}[tb]
\centering
\begin{tabular}{M{0.14\textwidth}M{0.14\textwidth}M{0.14\textwidth}}
(a) all hinges oriented in $y$: &  (b) oriented in $x$ and $y$: &  (c) oriented in tension:\\
\includegraphics[width=0.12\textwidth]{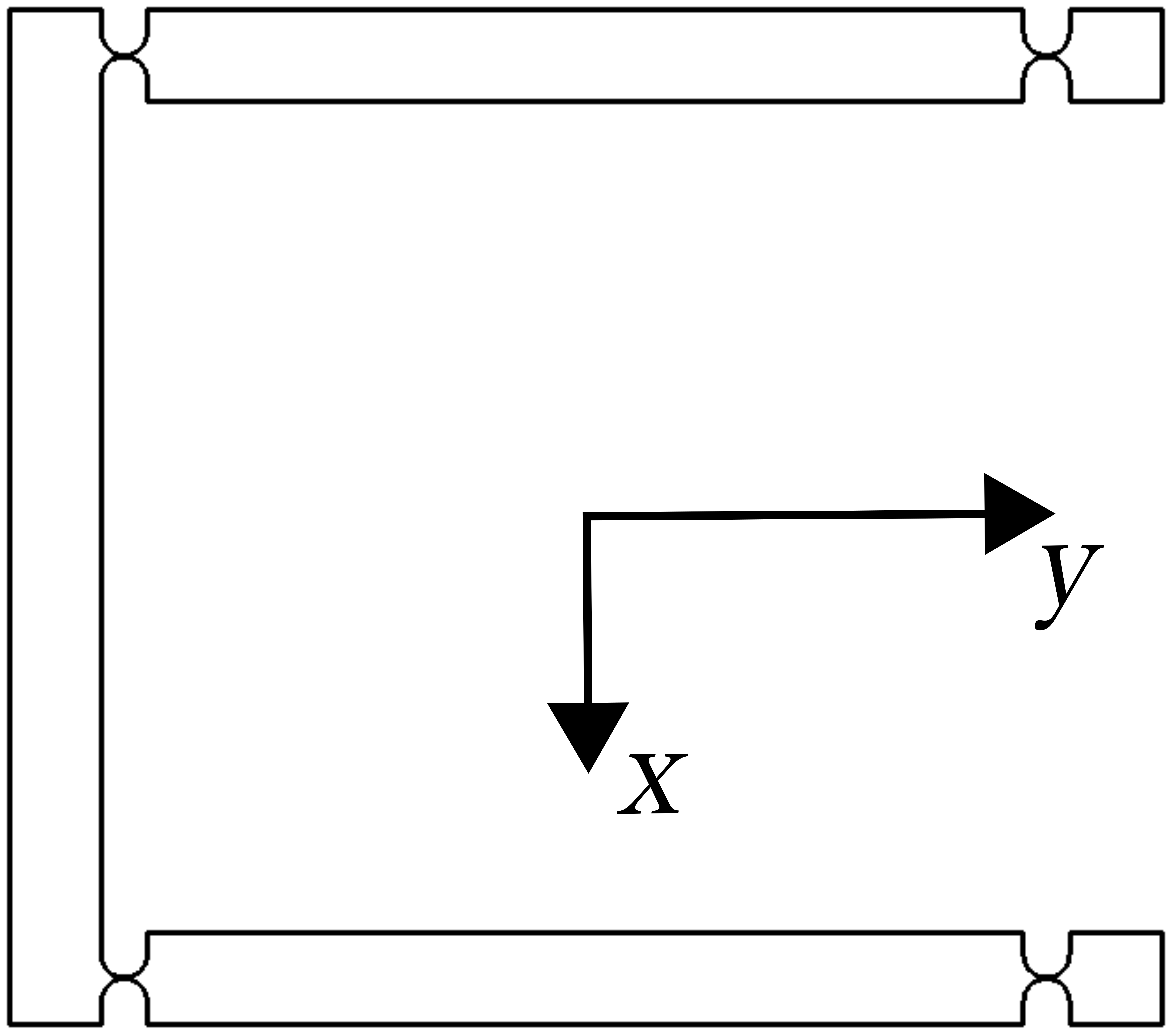} & 
\includegraphics[width=0.12\textwidth]{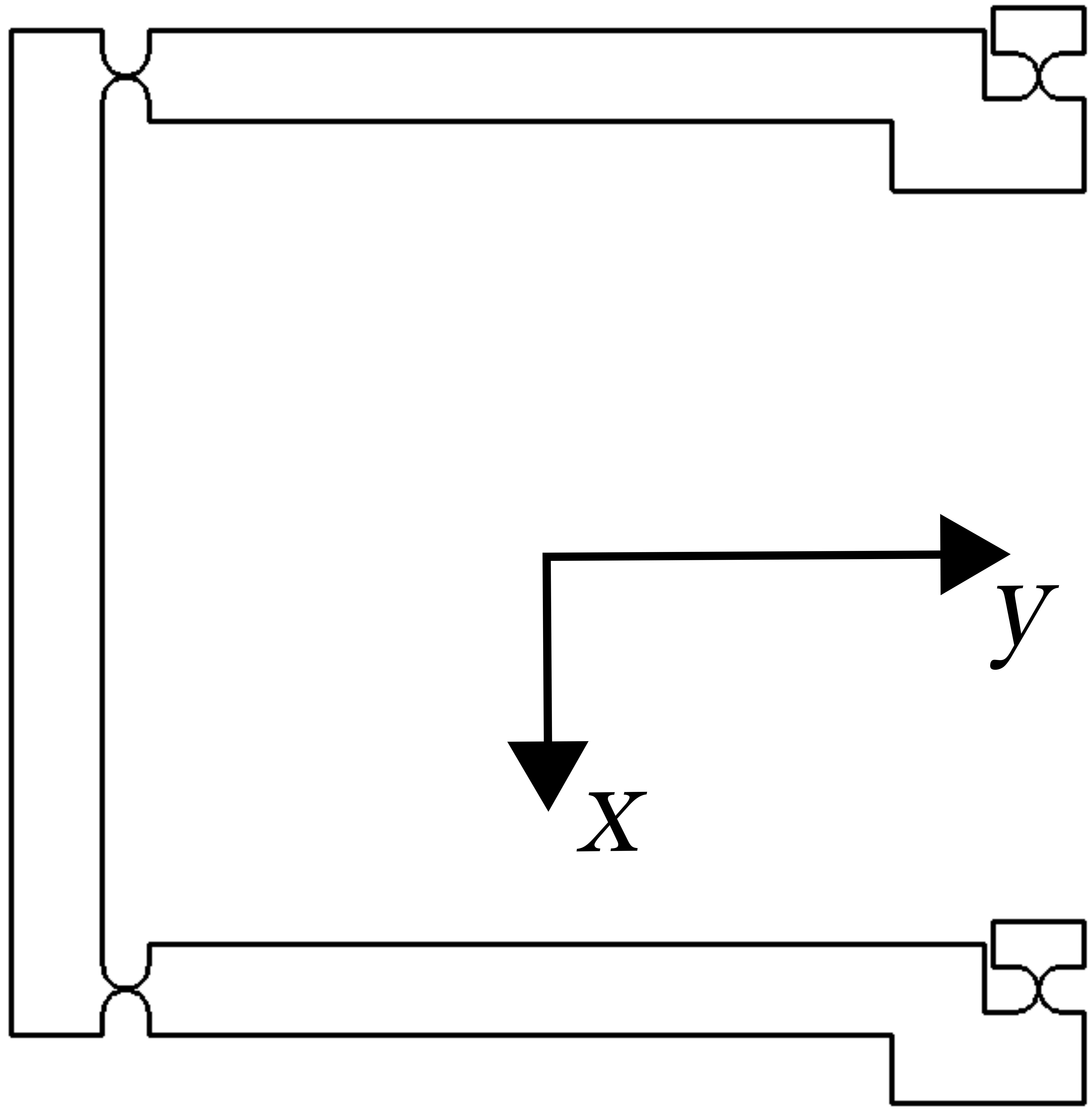} & 
\includegraphics[width=0.12\textwidth]{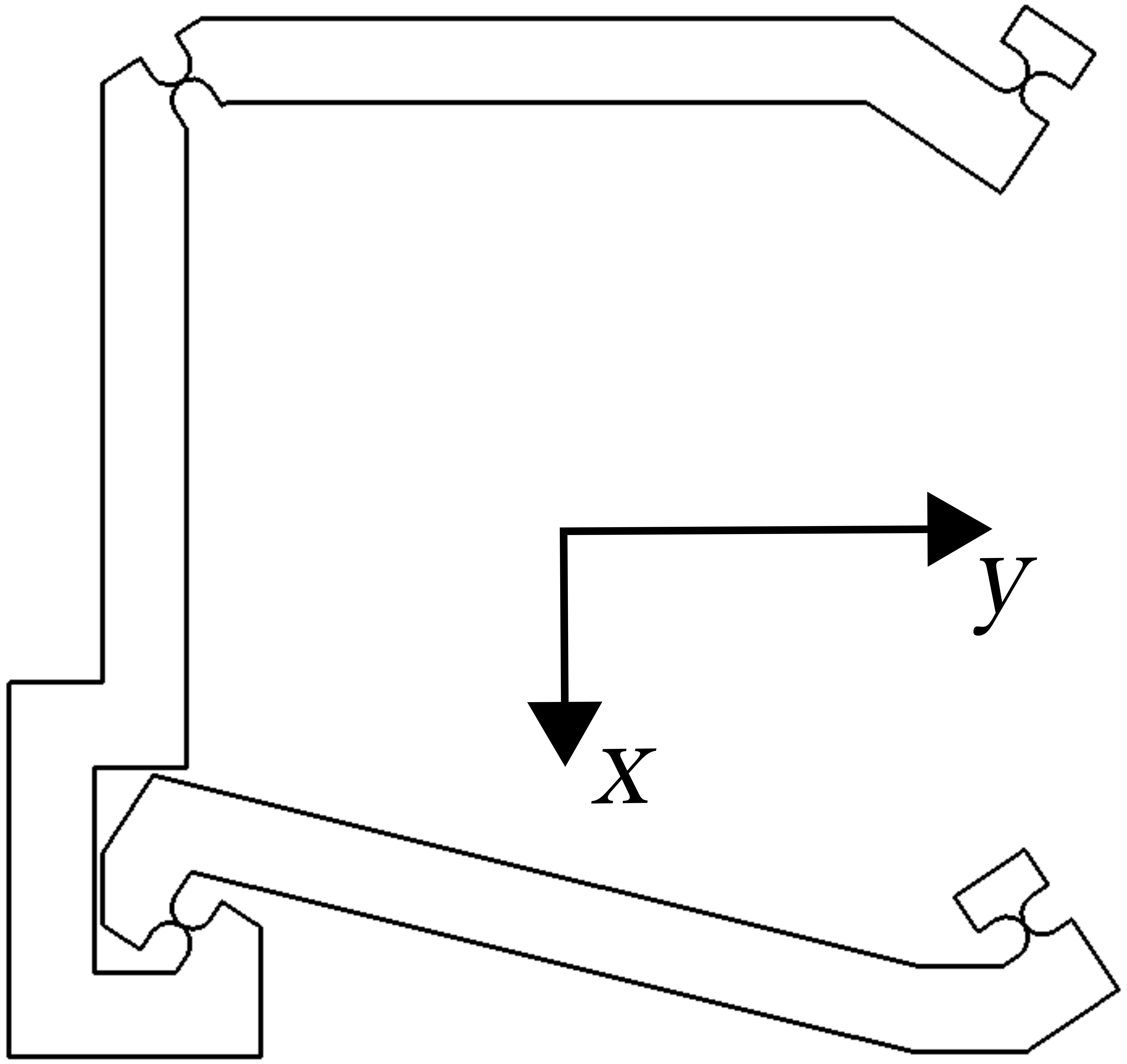}\\
\SI{110.5}{\mega\pascal} & 
\SI{80.9}{\mega\pascal} & 
\SI{9.0}{\mega\pascal}
\end{tabular}
  \caption{Maximum stress in the hinges as calculated by the finite element simulation  for three different hinge orientations. For these calculations an elastic modulus of $E=\SI{72}{\giga\pascal}$ and a Poisson's ratio of $\nu=0.33$ were used. The maximum stress arises in all configurations in the center of each back hinge.
  \label{fig:hingeorientations}}

\end{figure}

Three potential hinge orientations in the mechanism are investigated, see Fig.~\ref{fig:hingeorientations}.
In (a) all hinges are oriented along $y$, in (b) the two back hinges point along $x$ and the front hinges along $y$.
Finally, in (c) the hinges are oriented along the force vectors such that all hinges are in tension in the nominal zero position. For a mathematical derivation of the orientation of the hinge force vectors, see section~5 in~\cite{Keck.2020}.
For all three cases, a simplified design is studied using finite element simulation in ANSYS Workbench\footnote{Certain commercial equipment, instruments, and materials are identified in this paper in order to specify the experimental procedure adequately. Such identification is not intended to imply recommendation or endorsement by the National Institute of Standards and Technology, nor is it intended to imply that the materials or equipment identified are necessarily the best available for the purpose.}.   
Two analyses are performed to determine the stress in the hinges and stiffness of the mechanism.

For each geometry a finite element model is calculated in each analysis. Both simulations use quadratic elements and a nonlinear solver. 

For the stiffness analysis masses and the gravitational vector are excluded from the simulation. 
Hence, the only forces in the stiffness simulation are generated by the hinges.

In order to obtain the stiffness of the mechanism $K$, the coupler in the model is displaced vertically by \SI{0.1}{\milli\meter}, and the simulated restoring force is recorded.

The result of the three finite element calculations show that the stiffness is mostly independent of hinge orientation.
In each case, the result for the pure elastic part of the mechanism stiffness is $K\approx\SI{8.1}{\newton \per \meter}$. This result is remarkable.  The admixture of applied transversal force and torque on a given hinge under deflection changes with orientation and so does its stiffness leading one to expect a larger variation in stiffness. In this design, the effect is negligible. 
Similarly, adding the gravitational load to the hinges does not change their torsional stiffness significantly.

\begin{figure}[htb]
\centering
\includegraphics[width=1.0\columnwidth]{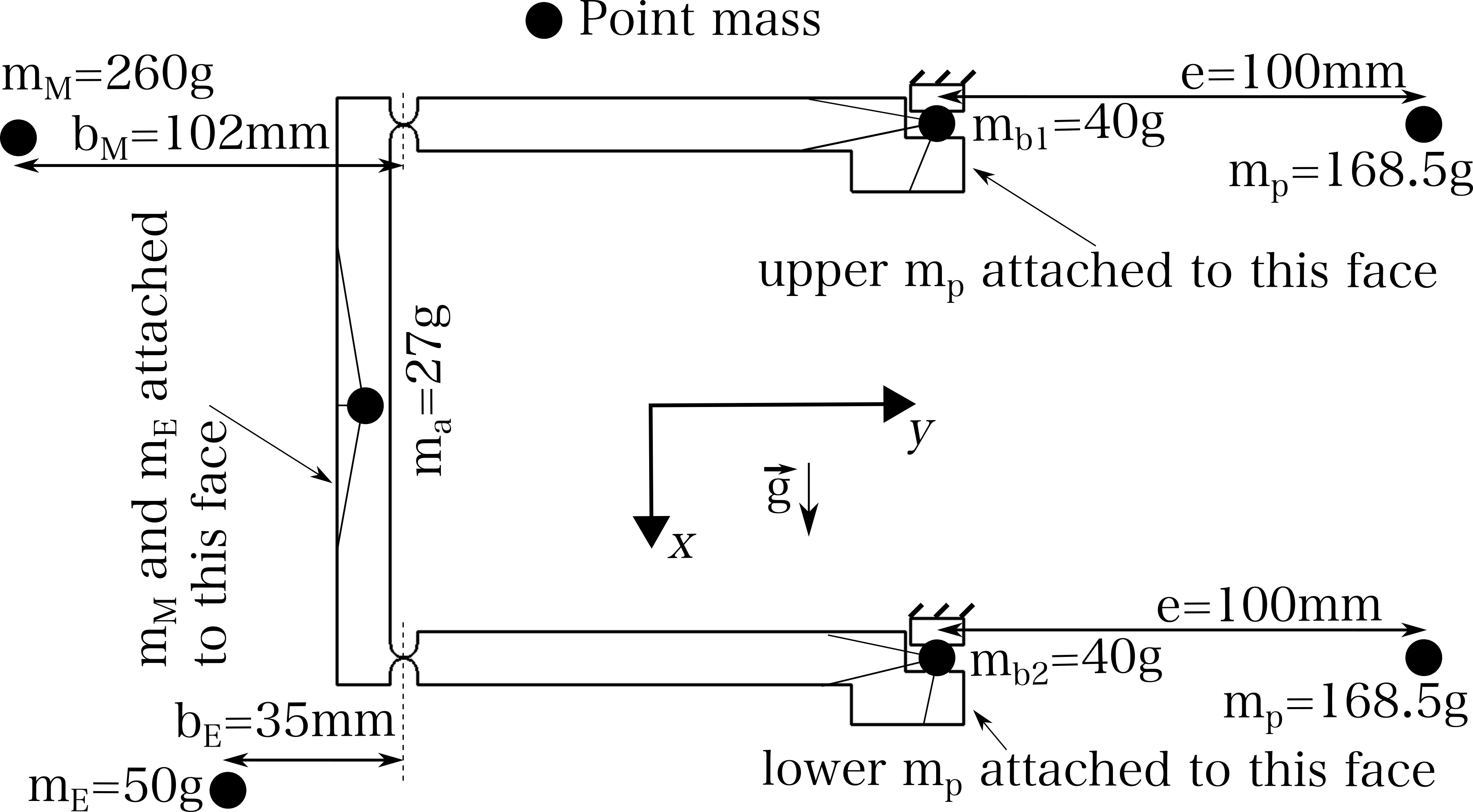}
\caption{Mass placement used for the hinge stress analysis via the finite element method. Variation (b) in Fig.~\ref{fig:hingeorientations} serves here as representation. The values for the masses and lengths are derived from preliminary designs and investigations of the functional components included in the design of the balance and are given in Table.~\ref{tab:parameters}.
\label{fig:nullsim}}
\end{figure}

For the hinge stress analysis, see Fig.~\ref{fig:nullsim}, the gravitational vector and masses are included in the simulation. All masses are modelled as points and the locations of the centers of mass of the swings coincide with the back pivots as assumed in the rigid body model. 
The mechanism is  considered with an equilibrium at the nominal zero position ($N_\mathrm{eq}=\SI{0}{}$) without applying further external displacements.
The result of the calculation is the maximum equivalent stress that occurs in the hinges of the mechanism.

In the hinge stress analysis, the differences in the calculated outcome is tremendous. 
For the two extreme cases it differs by more than an order of magnitude,  $\SI{110.5}{\mega \pascal}$ and $\SI{9.0}{\mega \pascal}$ for (a)  and (c) in Fig.~\ref{fig:hingeorientations}, respectively. Note that the maximum stress arises in all configurations in the center of each back pivot.

Orientation (c) in Fig.~\ref{fig:hingeorientations} clearly performs best. It has the smallest stress while being comparable in  elastic stiffness to the other orientations.
The result follows intuition, because here the hinges are loaded along the force vector, and the loading is in tension rather than compression. The latter can lead to buckling in these ultra thin notch flexures.

\subsection{Complete three dimensional model}

\begin{figure}[htb]
\begin{center}
\includegraphics[width=0.65\columnwidth]{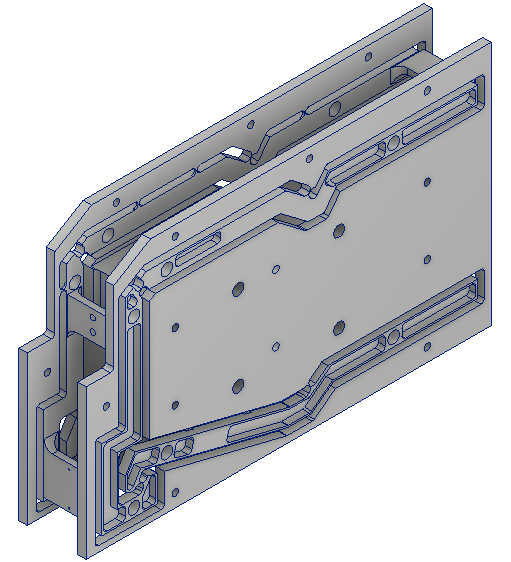}\\
\includegraphics[width=1.0\columnwidth]{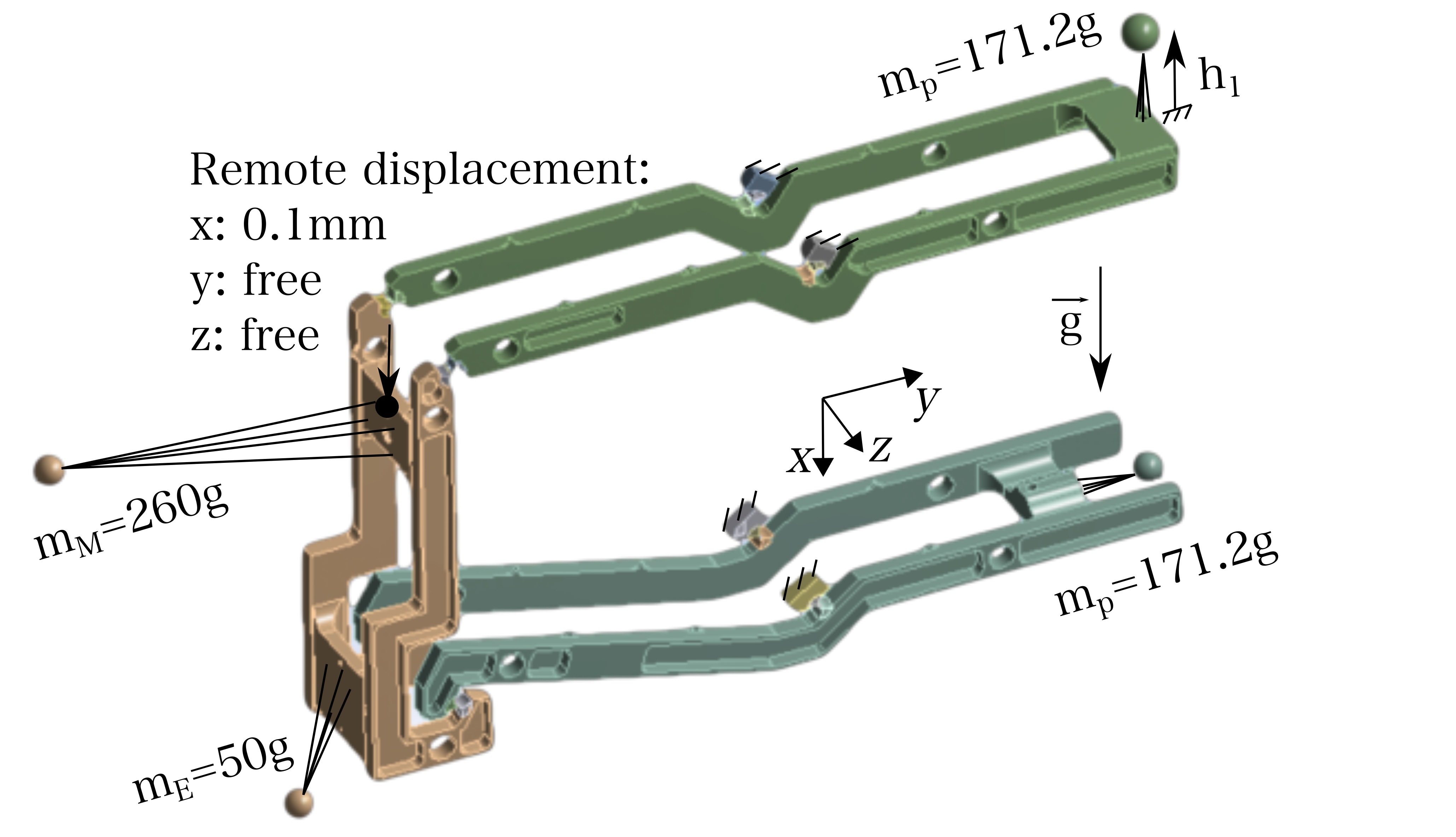}

\caption{The upper image shows the CAD model of the designed mechanism. It consists of two separate mechanisms with a connector in between to prevent corner loading and provide stability to the structure. The lower image shows the geometry of the mechanism illustrating the boundary conditions for finite element simulation.  Material density was considered with $\rho=\SI{2.8}{\gram \per \centi\meter^3}$ for aluminum. Note that the center of mass of each rotating link was designed to coincide with its pivot. Note also, that the weight of the coupler differs slightly from what was assumed in the first simulation in Fig.~\ref{fig:nullsim}. In the real CAD model it is $m_\mathrm{a}=\SI{32.4}{\gram}$ which causes the counterweights to have $m_\mathrm{p}=\SI{171.2}{\gram}$ each.\label{fig:femlast}}
\end{center}
\end{figure}

With the design choices described above a 3D CAD model is generated, see Fig.~\ref{fig:femlast}. 
The whole mechanism is built from an aluminum block measuring $\SI{241}{\milli \meter}$, $\SI{146}{\milli \meter}$, and $\SI{40}{\milli \meter}$ in length, height, and width. 

The moving part of the mechanism consists of two planar structures each $\SI{5}{\milli\meter}$ thick and spaced $\SI{30}{\milli\meter}$  apart. 
They are connected at four locations with connectors and move as one. 
The defining features in both planes are machined in a single fixtured position and are therefore nominally identical. 
The front and back plates protect the moving parts of the mechanism.
With the chosen approach, the attachments and counterweights can be mounted at the plane of symmetry between the two mounting plates. This configuration eliminates parasitic rotations about the y axis while preserving monolithic machinability.

The moving parts, the two swings and the coupler are separated from the plates by a $\SI{4}{\milli \meter}$ wide channel that is milled through both plates simultaneously. 
The channel runs not completely thru but is interrupted by the four hinges and sixteen sacrificial bridges. 
After milling, an  electrical wire discharge machine is used to precisely contour the hinges.
The bridges block the motion of the mechanism and provide sufficient stability for all machining steps, and are carefully removed at the end of the machining process.
The center of the mechanism is solid except for a few through holes, providing thermal and mechanical stability.
All four connectors have tapped holes. To the front connectors the mirror and capacitor can be mounted. Trim masses can be attached to the two back connectors, see section~\ref{sec:finaldesign}.
Two additional plates not shown in the drawing can be bolted on to the front and back plane. Six transportation safety pins through these plates immobilize the moving part of the mechanism during transport, see Fig.~\ref{fig:lowangleview}.

\section{Physical properties of the final design}
\label{sec:finaldesign}

With the final design, a second iteration of finite element analysis was performed with all the components necessary for laser power measurement. Expected values for the masses of the mirror, and the inner electrode were assigned. The two identical counterweights were chosen such that the balance is at the nominal zero position, i. e.,   $N_{\mathrm{eq}}=\SI{0}{}$,  see Eq.~\ref{eq:N_eq}. These four masses  are concentrated at points, while the masses of the coupler and the swings were assumed to be distributed, see Fig.~\ref{fig:femlast}.

The swing assembly consists of the upper/lower swing and the upper/lower counter mass plus half the mass to the left of the front flexures, see Eq.~\ref{eq:N_eq}. The center of mass of the lower swing assembly coincides vertically with the lower back pivot point and can be finally adjusted horizontally. The opposite is true for the upper swing assembly. Here, the center of mass coincides horizontally with the upper back pivot point and can be finally adjusted vertically. 

With the lower mass, the restoring torque of the balance is adjusted which allows to adjust the equilibrium position of the balance close to the nominal zero position of the linkage.  With the upper mass the stiffness of the balance is adjusted. If the center of mass of the upper swing assembly coincides with the pivot $(h_\mathrm{1}=0)$, the restoring torque is provided by the pivots alone. By moving the center of mass up, the restoring torque is reduced by the gravitational moment of the mass. The drawback of this method is that the gravitational moment changes with tilt of the balance frame. Hence, the tilt sensitivity increases as the mass of the swing assembly moves away from the pivot point.

\begin{figure}[htb]
\begin{center}
\includegraphics[width=1.0\columnwidth]{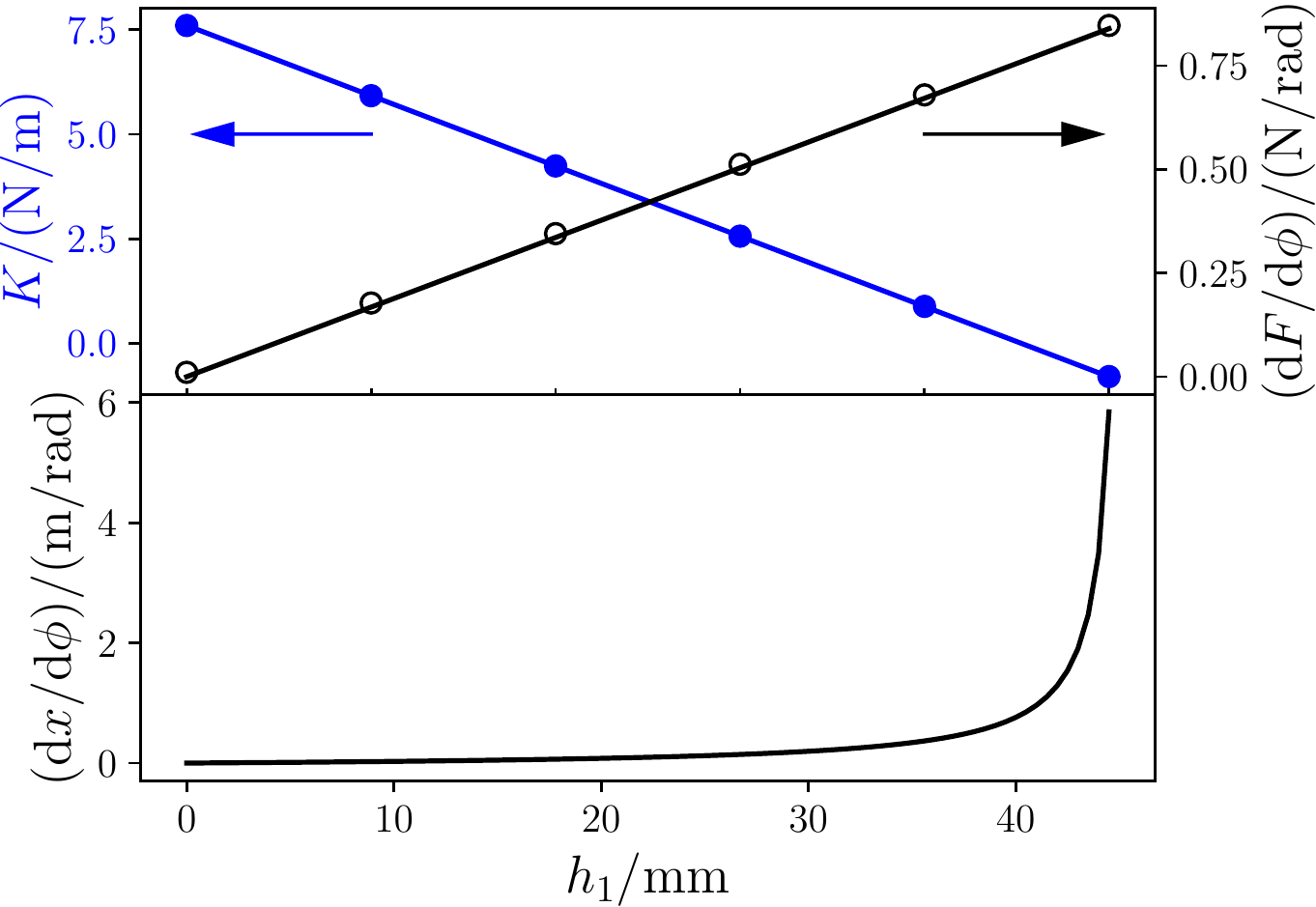}
\caption{The upper plot shows the analytical (solid lines) and finite element simulation (dots) results for the mechanism stiffness (blue filled circle) and the error force sensitivity due to ground tilt $\phi$ (black empty circle) for different compensation mass positions $h_\mathrm{1}$ in closed loop. The lower plot shows the sensitivity of the excursion of the coupler in $x$ due to ground tilt $\phi$ for different compensation mass positions $h_\mathrm{1}$ in open loop. Note that the excursion is measured with respect to the equilibrium in the nominal zero position of the coupler for $\phi=0$.\label{fig:tilt}}
\end{center}
\end{figure}

A finite element simulation is performed with the upper mass in six positions to investigate the trade-off between stiffness reduction and tilt sensitivity.
The results of these simulations and the calculation of the analytical model using Eqs.~\ref{eq:K_eq},~\ref{eq:transstiff} and~\ref{eq:K_eq,phi} are displayed in the upper plot in Fig.~\ref{fig:tilt}. 
The data in blue indicated by the left axis show the linear stiffness of the mechanism as a function of $h_\mathrm{1}$. 
Black data points with the scale on the right show  the produced torque on the coupler.
The circles are calculated with finite element simulation, and the lines from analytical equations. 
A good agreement indicates the validity of the analytical equations.

The goal is to obtain a linear stiffness of the coupler of $K\le\SI{0.1}{\newton \per \meter}$. Hence, $h_\mathrm{1}\ge\SI{44.12}{\milli\meter}$.

As indicated in the upper plot in Fig.~\ref{fig:tilt} a large $h_\mathrm{1}$  increased the tilt sensitivity. Fig.~\ref{fig:tilt} provides more detailed information. 
The plots show both the sensitivity of the balance readout to ground tilt in open (upper plot) and closed (lower plot) loop.
In open loop, the ground tilt causes a deflection of the coupler in vertical direction with respect to the metrology frame. 
In closed loop, a restoring force generated by the capacitor is necessary to maintain the coupler at the desired equilibrium position. 
In either case, a static tilt will drop out, since the external force is modulated. 
Hence only a tilt that occurs on the same time scale as the laser light modulation will contribute a bias to the measurement.  
We estimate such a modulation of the tilt is  $\SI{1}{\nano \radian}$.
Using $h_\mathrm{1}=\SI{44.12}{\milli\meter}$, a spurious force of $\SI{0.76}{\nano \newton}$ will be indicated in this case, which has no significant impact to the measurement result.

In addition to the above, the dynamic behaviour of the balance is also important to consider. 
The eigenfrequencies of the device are obtained with a modal analysis. 
A finite element simulation was carried out with the system at equilibrium,  $h_\mathrm{1}=\SI{44.12}{\milli\meter}$ and the mass distribution as shown in Fig.~\ref{fig:femlast}.
For the first resonance in the system, the oscillation along the force measurement axis ($x$), the eigenfrequency is  $f=\SI{0.51}{\hertz}$. 
The next resonance at $f=\SI{19}{\hertz}$ is the out of plane bending of the swings. 
A total of seven resonances occur below $\SI{100}{\hertz}$. 
Since the next lowest frequency is more than an order of magnitude away from the most compliant mode, the influence of the higher order modes on the measurement is believed to be negligible, or can easily be mitigated with appropriate filtering.

\begin{figure}[htb]
\begin{center}
\includegraphics[width=1.0\columnwidth]{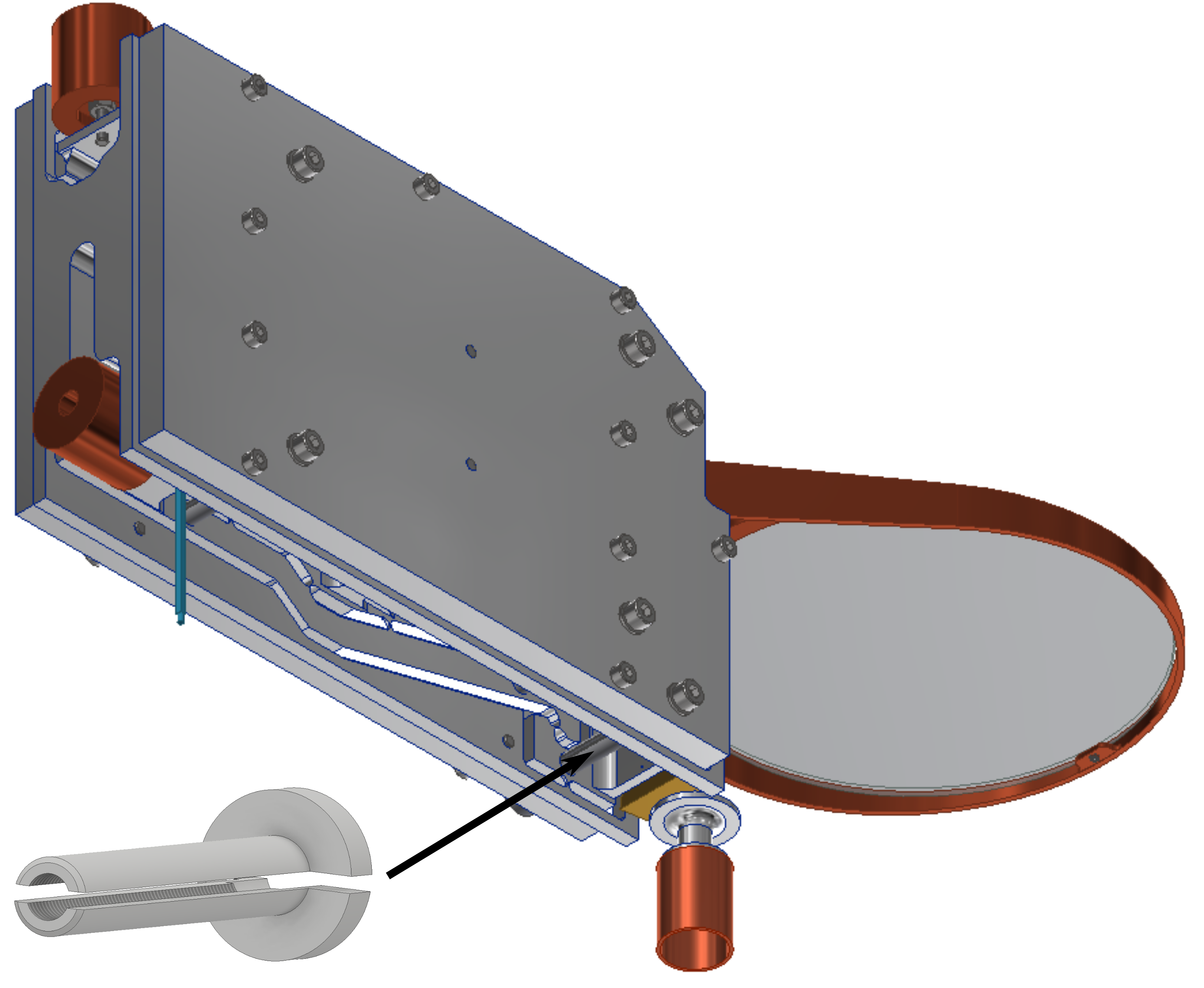}
\caption{Low angle view of the assembled mechanism. The inset on the bottom shows, not to scale, one of the total six transport safety pins.\label{fig:lowangleview}}
\end{center}
\end{figure}

A low angle view of the device with protective side plates installed is shown in Fig.~\ref{fig:lowangleview}. The side plates allow mounting, add thermal mass, and provide holes for transport safety pins, see inset in Fig.~\ref{fig:lowangleview}. 
Each safety pin is a spring sleeve. It can be compressed and inserted in the assembly without applying an insertion force. By engaging a screw in the sleeve, it expands and the mechanism is locked.
Both side plates were manufactured in one setup together with the mechanism, ensuring that the holes for the safety pins are precisely aligned. Hence, the pins lock the mechanism without additional forces and with zero clearance.

\section{Summary and Outlook}
An electrostatic force balance mechanism was developed to measure the photon pressuren force of a $\SI{100}{\kilo \watt}$ laser. 
The task required  the measurement of a $\SI{667}{\micro \newton}$ force with a relative uncertainty of $\SI{1e-3}{}$ in air.
Therefore, a portable monolithic parallelogram linkage was designed.  At first, the impact  of machining tolerances on the  corner loading error was investigated as a function of the two length in the parallelogram, $a$ and $b$. It was found, that corner loading is generally minimal when linkage dimensions are equal, $a=b$, and decreases with increasing $a+b$. The largest possible size in this application is $a=b=\SI{100}{\milli \meter}$ and was chosen for the final design.

Furthermore, the basic equation of motion was derived according to the Lagrange equations of the second kind and results were confirmed with finite element simulation. With these, the static and dynamic behavior of the system could be optimized by parameter variations. The most critical static properties of the mechanism are its equilibrium condition and the linear stiffness of the moving balance coupler.
Two moveable masses can be used to adjust both the equilibrium position and the linear stiffness of the coupler in the mechanism independently. In the final design  the stiffness of the mechanism in the measuring direction can be adjusted from $K=\SI{7.6}{\newton \per \meter}$ to below zero according to finite element simulation. 
Lowering the stiffness increases the sensitivity of the balance to ground tilt. Analyzing this trade-off with both finite element simulation and analytical modeling allowed to estimate this impact, which was found to be negligible for the measurement considering a desired value for the mechanism stiffness at the coupler of $K=\SI{0.1}{\newton \per \meter}$.

After manufacturing and system integration, experiments will be required to study especially the effects of internal heating to the measurement readout due to absorbed laser light from the \SI{100}{\kilo \watt} laser. Also, air currents on the mirror due to the operation of the balance are expected to cause a significant amount of noise. 
In order to mitigate these problems, heat and draft shields will be installed in the final setup.

Mechanically,  hysteresis due to anelastic after-effects in the flexures might cause time-dependent restoring forces, which would bias the measurement. Hysteresis is difficult to study theoretically due to limited information provided by literature, but for now, these anelastic forces were not considered to be problematic, because the maximum stress in the hinges is within ranges suggested by Sydenham~\cite{.k} to keep anelastic effects small.
Further experimental work will be carried out to verify this.

In conclusion, the mechanism described above fulfills the criteria necessary for measurement of photon pressure force from high power laser systems, with relative uncertainties  below \SI{1e-3}{}.

\section{Acknowledgements}
This work has been done in close cooperation between the Physical Measurement Laboratory at NIST and the Precision Engineering Group at Technische Universit\"at Ilmenau. We also want to thank the NIST internal reviewers Vincent Lee and David Newell for constructive feedback.

\section*{Appendix A: Summary of the parameters used for the final design}

The numerical values for the important mechanical parameters of the final design are shown in Table~\ref{tab:parameters}. The locations of the parameters are indicated in Fig.~\ref{fig:mechanismvertical}.

\begin{table}[ht]
    \centering
    \begin{tabular}{l  S[table-format=3.1] p{0.6\columnwidth}}
        \toprule
        &&\multicolumn{1}{l}{\textbf{masses}}    \\
        \hline
        $m_\mathrm{a}$ & \SI{32.4}{\gram} & coupler \\
        $m_\mathrm{E}$ & \SI{50.0}{\gram} & inner capacitor electrode \\
        $m_\mathrm{M}$ & \SI{260.0}{\gram} &  mirror including  holder \\
        $m_\mathrm{p}$ & \SI{171.2}{\gram} & each counter weight \\
        \\
        &&\multicolumn{1}{l}{\textbf{lengths}}    \\
        \hline
        $a$ & \SI{100.0}{\milli \meter} &  coupler \\
        $b$ & \SI{100.0}{\milli \meter} &  each swing \\
        $e$ & \SI{100.0}{\milli \meter} &  each compensation arm \\
        \\
                &&\multicolumn{1}{l}{\textbf{horizontal distances between}}    \\
        \hline
        $b_\mathrm{M}$ & \SI{102.0}{\milli \meter} &  coupler \& mirror center of mass \\
        $b_\mathrm{E}$ & \SI{35.0}{\milli \meter} &  coupler \& electrode center of mass \\ 
        $\Tilde{L}$ & \SI{89.0}{\milli \meter} &  external force \& compensation force \\
        \\
        &&\multicolumn{1}{l}{\textbf{vertical distances between}}\\
        \hline
        $h_\mathrm{1}$ & \SI{44.1}{\milli \meter} & upper $m_\mathrm{p}$ \& pivot  for $K=\SI{0.1}{\newton \per \meter}$ \\
        \bottomrule
    \end{tabular}
    \caption{Overview of mechanical relevant parameters for the final design. The horizontal distances to the coupler are measured to its center.}
    \label{tab:parameters}
\end{table}

\section*{Appendix B: Derivation of the equation of motion}

\label{section:lagrange}
The equation of motion can be derived using the Lagrange equations of the second kind.
The Lagrangian is given by
\begin{equation}
    \mathcal{L}(q_i,\dot{q_i} , t) = T - U,\label{eq:lagrangian}
\end{equation}
where  $q_\mathrm{i}$  are the generalized coordinates, $i=1...n$ for $n$ degrees of freedom.\\
The parallelogram linkage has one degree of freedom, $q_1 = \Theta$. This generalized coordinate can be written as the sum of the tilt of metrology frame, $\phi$, and the deflection of the balance with respect to the metrology frame, $\gamma$, i. e., $\Theta = \phi + \gamma$.

The position of nine point masses, as shown in Fig.~\ref{fig:mechanismvertical} can be written as products of lever arms and sines and cosines of the corresponding angles. They are,

\begin{equation}
    \vv{r}_{\mathrm{mb1}} = \left( \begin{array}{c} 
    -\frac{\ds a} {\ds 2} \cos{\phi}\\[2ex]
    -\frac{\ds a}{\ds 2} \sin{\phi}
    \end{array}\right), \\[3ex]
\end{equation}

\begin{equation}
    \vv{r}_{\mathrm{mb2}} = \left( \begin{array}{c} 
    \frac{\ds a} {\ds 2} \cos{\phi}\\[2ex]
    \frac{\ds a}{\ds 2} \sin{\phi}
    \end{array}\right), \\[3ex]
\end{equation}

\begin{equation}
    \vv{r}_{\mathrm{ma}} = \left( \begin{array}{c} 
    b \sin{(\phi + \gamma)}\\[2ex]
    - b \cos{(\phi + \gamma)}
    \end{array}\right), \\[3ex]
\end{equation}

\begin{equation}
     \vv{r}_{\mathrm{mp1}} = \left( \begin{array}{c} 
    -\frac{\ds a} {\ds 2} \cos{\phi} -e \sin{(\phi + \gamma)}\\[2ex]
    -\frac{\ds a}{\ds 2} \sin{\phi} + e \cos{(\phi + \gamma)}
    \end{array}\right), \\[3ex]
\end{equation}

\begin{equation}
     \vv{r}_{\mathrm{mp2}} = \left( \begin{array}{c} 
    \frac{\ds a} {\ds 2} \cos{\phi} -e \sin{(\phi + \gamma)}\\[2ex]
    \frac{\ds a}{\ds 2} \sin{\phi} + e \cos{(\phi + \gamma)}
    \end{array}\right), \\[3ex]
\end{equation}

\begin{equation}
    \vv{r}_{\mathrm{mh1}} = \left( \begin{array}{c} 
    -\frac{\ds a} {\ds 2} \cos{\phi} -h_1 \cos{(\phi + \gamma)}\\[2ex]
    -\frac{\ds a}{\ds 2} \sin{\phi} - h_1 \sin{(\phi + \gamma)}
    \end{array}\right), \\[3ex]
\end{equation}

\begin{equation}
    \vv{r}_{\mathrm{mh2}} = \left( \begin{array}{c} 
    \frac{\ds a} {\ds 2} \cos{\phi} -h_2 \cos{(\phi + \gamma)}\\[2ex]
    \frac{\ds a}{\ds 2} \sin{\phi} - h_2 \sin{(\phi + \gamma)}
    \end{array}\right), \\[3ex]
\end{equation}

\begin{equation}
    \vv{r}_{\mathrm{mM}} = \vv{r}_{\mathrm{ma}} + \left( \begin{array}{c} 
    -\frac{\ds a_{\mathrm{M}}} {\ds 2} \cos{\phi} +b_{\mathrm{M}} \sin{(\phi +     \gamma)}\\[2ex]
    -\frac{\ds a_{\mathrm{M}}}{\ds 2} \sin{\phi} - b_{\mathrm{M}} \cos{(\phi + \gamma)}
    \end{array}\right), \; \mbox{and} \\[3ex]
\end{equation}

\begin{equation}
    \vv{r}_{\mathrm{mE}} = \vv{r}_{\mathrm{ma}} + \left( \begin{array}{c} 
    \frac{\ds a_{\mathrm{E}}} {\ds 2} \cos{\phi} +b_{\mathrm{E}} \sin{(\phi +     \gamma)}\\[2ex]
    \frac{\ds a_{\mathrm{E}}}{\ds 2} \sin{\phi} - b_{\mathrm{E}} \cos{(\phi + \gamma)}
    \end{array}\right).
\end{equation}

Note, the first/second line of the vectors indicate the $x\mbox{-}/y$-coordinate, with the $x$ being positive in the downward vertical direction.

The total kinetic energy is the sum of the translational ($_\mathrm{t}$) and rotational ($_\mathrm{r}$) energies, $T = T_{\mathrm{t}} + T_{\mathrm{r}}$.

The kinetic energy of the two swings can be captured by a single rotational term. The centers of rotation are pivots A$_\mathrm{0}$ and B$_\mathrm{0}$, respectively. The moment of inertia is calculated around these centers of rotations considering the bars as rods with length $b+e$. It is
\begin{equation}
    J_\mathrm{b} = \frac{m_\mathrm{b}}{12} \left(b+e \right)^{2}.
\end{equation}
The motion of the coupler is described as a translation of its center of mass with velocity $\dot{\vv{r}}_\mathrm{ma}$ and a rotation about $\phi$ around its center of mass. The moment of inertia of the coupler $J_{_\mathrm{a}}$ is
\begin{equation}
    J_{\mathrm{a}} = \frac{m_\mathrm{a}}{12} a^2.
\end{equation}
Hence the sum of the translational and rotational kinetic energies are
\begin{multline}
    T_{\mathrm{t}} = \frac{1}{2}   m_\mathrm{a}   \dot{\vv{r}}_\mathrm{ma}^2 + \frac{1}{2}   m_\mathrm{p}   \dot{\vv{r}}_\mathrm{mp1}^2 + \frac{1}{2}   m_\mathrm{p}   \dot{\vv{r}}_\mathrm{mp2}^2 \\
    \hspace{6ex}+ \frac{1}{2}   m_\mathrm{h}   \dot{\vv{r}}_\mathrm{mh1}^2 + \frac{1}{2}   m_\mathrm{h}   \dot{\vv{r}}_\mathrm{mh2}^2 \hfill\\ 
    \hspace{6ex}+\frac{1}{2}   m_\mathrm{M}   \dot{\vv{r}}_\mathrm{mM}^2 + \frac{1}{2}   m_\mathrm{E}   \dot{\vv{r}}_\mathrm{mE}^2, \; \mbox{and} \hfill
\end{multline}
\begin{flalign}
     \hspace{2.5ex}&T_{\mathrm{r}} = \frac{1}{2} J_{\mathrm{a}}   \dot{\phi}^2 + J_{\mathrm{b}}   (\dot{\gamma} + \dot{\phi})^2.& \hfill
\end{flalign}

The potential energy of the system is a sum of the energy stored in the torsional stiff pivots  ($_\mathrm{k}$) and the sum of the gravitational energies of the  masses  ($_\mathrm{m}$), $U = U_{\mathrm{k}} + U_{\mathrm{m}}$.
It is
\begin{flalign}
    \hspace{2.5ex}&U_{\mathrm{k}} = 4 \frac{1}{2}   \kappa_\mathrm{s}   \gamma^2, \;\mbox{and}& \hfill
\end{flalign}
\begin{multline}
    U_{\mathrm{m}} = 
    -g \, (m_{\mathrm{b}}   r_\mathrm{mb1x} 
    +\,m_{\mathrm{b}}   r_\mathrm{mb2x} 
    +\, m_{\mathrm{a}}   r_\mathrm{max} \\
    \hspace{6ex}+\, m_{\mathrm{p}}   r_\mathrm{mp1x} 
    +\, m_{\mathrm{p}}   r_\mathrm{mp2x}    
    +\,m_{\mathrm{h}}   r_\mathrm{mh1x}  \hfill\\
    \hspace{6ex}+\,m_{\mathrm{h}}   r_\mathrm{mh2x} 
    +\,m_{\mathrm{M}}   r_\mathrm{mMx}  
    +\, m_{\mathrm{E}}   r_\mathrm{mEx}). \hfill
\end{multline}
Consistent with the main text, the torsional stiffness of a single flexure hinge is given by $ \kappa_\mathrm{s}$.
Only one external moment needs to be considered. It arises from the stiffness adjustment spring and is
\begin{equation}
    Q_{\mathrm{e}} = -F_{\mathrm{F}} \Big(  \left(l_{\mathrm{1}}+b \cos{\gamma}\right)  \sin{\sigma}  + b \sin{\gamma} \cos{\sigma} \Big),
\end{equation}
 where $\sigma$ is the angle between the orthogonal of the metrology frame and the spring force $F_\mathrm{F}$, i. e., 
\begin{eqnarray}
    \sigma &=& \frac{\ds b \sin{\gamma}}{\ds l_{\mathrm{1}} + b \cos{\gamma}}, \;\mbox{and}\;\\
    F_{\mathrm{F}} &=& -(\lambda_1 - \lambda_0) k_\mathrm{b},\;\mbox{with}\\
    \lambda_{\mathrm{1}} &=& \sqrt{(b \sin{\gamma})^2 + (l_{\mathrm{1}} + b \cos{\gamma})^2}.
\end{eqnarray}
With the Lagrangian in Eq.~\ref{eq:lagrangian} and the previous considerations the equation of motion due to $\gamma$ yields
\begin{equation}
    \frac{\mbox{d}}{\mbox{d}t}\left(\frac{\partial \mathcal{L}}{\partial \dot{\gamma}}\right)-\frac{\partial \mathcal{L}}{\partial \gamma} =  Q_{\mathrm{e}}.
    \label{eq:L_with_Q}
\end{equation}

Taking the derivatives and regrouping the expressions yields a compact result,

\begin{equation}
    J \ddot{\gamma} + \kappa \gamma = -J_{\mathrm{\phi}} \ddot{\phi}  + \kappa_{\mathrm{\phi}} \phi - N_{\mathrm{eq}},\label{eqA:eqm}
\end{equation}

where the coefficients are given by
\begin{multline}
    J = b^{2} m_\mathrm{E} + b^{2} m_\mathrm{M} + b^{2} m_\mathrm{a} + \frac{b^{2} m_\mathrm{b}}{6} + \frac{b e m_\mathrm{b}}{3} + \frac{e^{2} m_\mathrm{b}}{6} \\
    \hspace{6ex}  +2 e^{2} m_\mathrm{p} + h_\mathrm{1}^{2} m_\mathrm{h} + h_\mathrm{2}^{2} m_\mathrm{h}, \hfill \label{eqA:J_eq}
\end{multline}

\begin{multline}
J_{\mathrm{\phi}} = J +\frac{a h_\mathrm{1} m_\mathrm{h}}{2} - \frac{a h_\mathrm{2} m_\mathrm{h}}{2} + b b_\mathrm{e} m_\mathrm{E} + b b_\mathrm{m} m_\mathrm{M}  \hfill 
\end{multline}

\begin{multline}
    \kappa = 4 \kappa_\mathrm{s} - g m_\mathrm{h} \left(h_\mathrm{1} + h_\mathrm{2}\right) \\
    \hspace{6ex} - l_{\mathrm{1}} b k_\mathrm{b} \left(1 - \frac{\lambda_{\mathrm{0}}}{b+l_{\mathrm{1}}}\right), \hfill \label{eqA:K_eq}
\end{multline}
\begin{flalign}
     \hspace{2.5ex}&\kappa_{\mathrm{\phi}} = g m_{h} \left(h_\mathrm{1} + h_\mathrm{2}\right) \label{eqA:K_eq,phi},\;\mbox{and}& \hfill
\end{flalign}
\begin{flalign}
    \hspace{2.5ex}&N_{\mathrm{eq}} = g \left(- b m_\mathrm{E} - b m_\mathrm{M} - b m_\mathrm{a} + 2 e m_\mathrm{p}\right).& \hfill \label{eqA:N_eq}
\end{flalign}

\nocite{*}
\bibliography{references}

\end{document}